\documentclass[aps,prl,twocolumn,superscriptaddress]{revtex4-1}
\usepackage{amsmath,amssymb,amsfonts}
\usepackage[most]{tcolorbox}  

\usepackage[dvipsnames]{xcolor}
\usepackage{comment}
\usepackage[overload]{empheq}
\usepackage{upgreek}
\usepackage{lmodern,pdftexcmds}
\usepackage{hyperref}
\usepackage{lineno}
\usepackage[margin=1in]{geometry}

\usepackage{graphicx}
\usepackage{float}
\newcommand{\radboud}{Department of Artificial Intelligence, Donders Center for Cognition, Radboud University, Nijmegen, The Netherlands}

\newcommand{\tauphys}{School of Physics and Astronomy and the Center for Physics and Chemistry of Living Systems, Tel Aviv University, Tel Aviv 6997801, Israel}

\begin{document}
\graphicspath{{figures/}}
\title{Dynamical Spreading and Memory Retention of Particle Suspensions under
    Power Law Potential}
\author{Ido Fanto}
\email{idofanto@mail.tau.ac.il}
\affiliation{\tauphys}

\author{Yuval Rosenblum }
\affiliation{\tauphys}

\author{Ori Harel}
\affiliation{\tauphys}

\author{Matan Yah Ben Zion}
\affiliation{\radboud}

\author{Naomi Oppenheimer}
\email{naomiop@gmail.com}
\affiliation{\tauphys}

\date{\today}

\begin{abstract}
We study the overdamped dynamic spreading of a suspension of particles under a repulsive power law potential.  We predict that the suspension spreads in a self-similar form, with its radius growing in time with a power independent of the dimension. We confirm this prediction experimentally using magnetized colloids with dipolar repulsion. Numerical simulations corroborate the experiments and further predict a categorically different behavior at a critical power, below which the initial distribution is no longer concentrated at the origin. Instead, particles accumulate at the perimeter and retain a long-lived memory of their original pattern. Below this threshold, the initial distribution seeds the resulting pattern, encoding the future structure of a dynamically evolving system.

\end{abstract}

\maketitle

Power law potentials influence many phenomena governing our lives \cite{french2010long}. They are the reason the Earth orbits the sun and what causes our hair to stand on end on a dry winter day; they are how charged particles interact with each other and why a compass points to the north \cite{purcell2013electricity}. As such, it is no wonder that power law potentials have been the subject of numerous studies. Usually, the focus is on equilibrium configurations \cite{leble2017large,Agarwal2019Harmonic, dandekar2023dynamical, irvine2010pleats, guerra2018freezing}, or on phase transitions \cite{zahn2000dynamic, dillmann2013two}. Far less attention has been given to their dynamics \cite{ispolatov1996annihilation, ginzburg1997self, marcos2017formation, flack2023out,HelenaMassanaCid2019,PhysRevE.89.012306,EbertMaretKeim}, usually with respect to relaxation to equilibrium \cite{bouchet2010thermodynamics}. However,  both living and engineered colloidal systems are hardly ever static --- they grow and evolve out of equilibrium. 

In this Letter, we analyze the deterministic dynamics of an initially concentrated suspension of particles as it spreads in free space due to power law repulsive potentials, $U \propto 1/r^k$ (known as Riesz gases \cite{lewin2022coulomb,serfaty2024lecturescoulombrieszgases} for Newtonian particles). Here, we assume that the system is overdamped. The exponent, $k$, and the dimension, $d$, classify the interaction: for $k>d$ the interaction is short-ranged, as particles interact with their nearest neighbors. Common examples are magnetic dipoles \cite{baranov2008theoretical, landau2013electrodynamics}, or Rydberg interactions \cite{gallagher1988rydberg}; for $k\leq d$, the interaction is long-ranged and spans the system size~\cite{binney1992theory, campa2009statistical}, as in, e.g.,  plasmas \cite{nicholson1983introduction}, superfluids \cite{kosterlitz2018ordering}, and cold atoms \cite{zhang2017observation}, and is commonly used in random matrices \cite{dyson1962brownian}. As we will show, the dynamical evolution of the density profile is captured by three regimes: (i) origin centered for $k> d-2$, (ii) constant for $k = d-2$, and (iii) boundary centered for $k<d-2$ (resembling the coffee-ring effect \cite{deegan1997capillary}). These regimes exhibit vastly different relaxation times when two suspensions collide. Namely, for strongly long-ranged potentials with $k<d-2$ we observe a power law retention of the initial conditions, relaxing extremely slowly toward an isotropic absorbing state. Such memory retention and slow relaxation were observed in other systems of long-range interactions, such as in sheared suspensions \cite{keim2011generic, keim2019memory}, crumpled paper \cite{shohat2023logarithmic}, or globally coupled spin systems \cite{mukamel2005breaking, gupta2010slow}.
\begin{figure*}[htbp!]
\centering
\includegraphics[width = 0.9\linewidth]{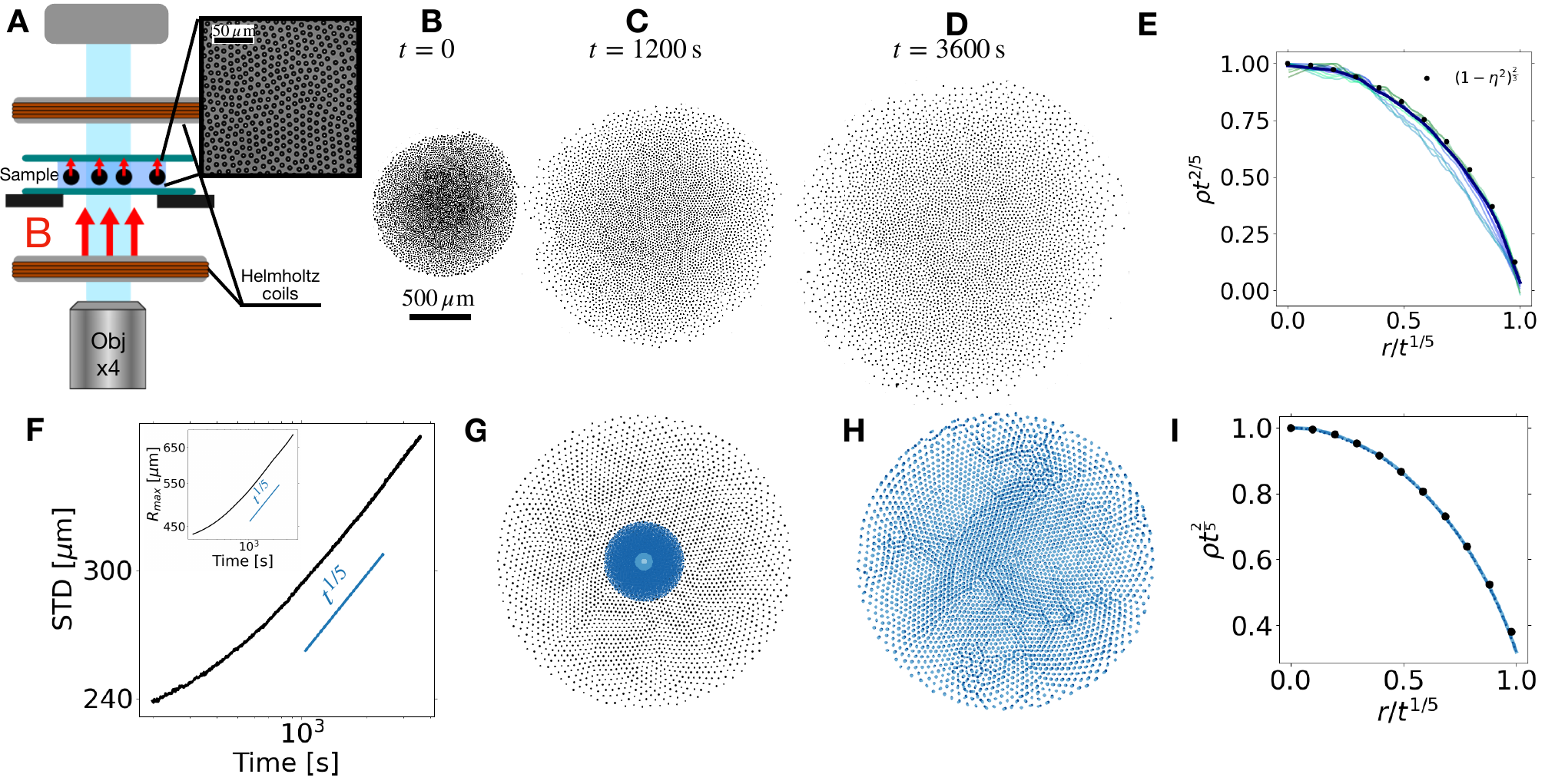}
\caption{Experiments and simulations of particles spreading under a $1/r^3$ pair-potential. (A) Schematics of the experimental setup. The inset shows the sample with a $\times20$ magnification. Triangular organization of the particles can be seen.  (B-D) Snapshots at three different times with $\times 4$ magnification: (B) Initially, (C) after 20 minutes, and (D) after one hour. (E) Density from four different experiments, at three different times. Rescaled with respect to $t^{2/5}$ as a function of the self-similarity parameter $r/t^{1/5}$. Rigid dark blue line is an average of all plots, and black dots are the theoretical prediction, faded  lines of the same color scheme correspond to the same experiment at different times. (F) The STD as a function of time showing ${\rm STD} \sim t^{1/5}$. Inset shows the maximal radius $R_{\rm max}\sim t^{1/5}$, averaged over the 50 most distant particles from the center of mass. (G-I) Results from molecular dynamics simulations of 3000 particles with a $1/r^3$ potential. (G)  Snapshots of the simulation at three different times ($t = 0.002, 10, 4800$). (H) The same snapshots from but rescaled according to Eq.~\ref{self-similar ansatz}. (I) Rescaled angular averaged density of the same densities showing a collapse to a single curve and a fit to the theoretical prediction given by Eq.~\ref{f in thesis}. }
\label{fig1}
\end{figure*}

In what follows, we derive the scaling dynamics of the suspension and show that since power law interactions lack a typical length or time scale, the density has a self-similar solution of the form \cite{PhysRevLett.52.1669}
\begin{equation}
\label{self-similar ansatz}
\rho(r,t) = At^{\gamma}f\left(\frac{Br}{t^{\beta}}\right) \ ; \ \gamma = -\beta d \ ; \ \beta = \frac{1}{k+2},
\end{equation}
More details on $A$ and $B$ are provided in End Matter. The exponent $\beta$ controls the density distribution and is independent of spatial dimension, which only affects the density amplitude.
We will verify this result by observing experimentally the two-dimensional (2D) spreading of perpendicularly magnetized colloids interacting by dipole-dipole repulsion scaling as $1/r^3$ \cite{zahn1999two, zahn2000dynamic}. We observe that the radius grows as $t^{1/5}$ as predicted by Eq.~\ref{self-similar ansatz}. The density profiles at different times collapse to a single curve given by Eq.~\ref{self-similar ansatz}. We further verify our results using molecular dynamics simulations of thousands of particles.
We go on to study the density profiles for different power laws in 2D (1D in End Matter).

We focus on systems in the overdamped regime, such that velocity is directly proportional to force through the mobility $\mu$, $\textbf{v} = \mu {\textbf{F}}  \sim 1/r^{k+1}$. The velocity of each particle is given by the force acting on it from all other particles. For a system of $N$ particles, the velocity of the $i^{\rm th}$ particle is given by
\begin{equation}
\label{discrete velocity}
\mathbf{v_{i}(r_{i})}=v_0 l^{k+1}\sum_{j\neq i}^{N}\left|\mathbf{r_{i}-r_{j}} \right|^{-(k+1)}\mathbf{\hat{r}_{ij}},
\end{equation}
with $v_0 = \mu F_0$ being the typical magnitude of the velocity, $F_0$ the typical magnitude of the force, and $l$ is a dimensionality constant.  

\textbf{Experiments.}
To test the collective dynamics experimentally, we used superparamagnetic particles of varying size and composition---$10\, \mu$m polystyrene (PS, density $1.5-2 \,{\rm g/cm^3}$) as seen in Fig.~\ref{fig1}, and $5\, \mu$m silicon dioxide ($\mathrm{SiO_{2}}$, density $1.6-1.8 \,{\rm gr/cm^3}$) in Fig.~\ref{fig:Sio expriment} in End Matter, both embedded with nanoparticles of iron oxide---we estimate the volumetric susceptibility to be between $ 0.04\leq\chi \leq0.4$. The dense particles are charge-stabilized and suspended in deionized water with $0.1\%w/v$ of Pluronic F-127 to improve particle stability, forming a quasi-2D arrangement at the bottom. Particles are then magnetized perpendicularly to the plane using Helmholtz coils (see Fig.~\ref{fig1}A). The induced dipole-dipole magnetic repulsion is of the form $U(r) \propto 1/r^{3}$. More precisely, for each two particles
$    U = \mu_0 r^{-3}\left[\textbf{m}_1\cdot \textbf{m}_2 - 3\left(\textbf{m}_1\cdot\hat{\textbf{r}}\right)
    \left(\textbf{m}_2\cdot\hat{\textbf{r}}\right)
    \right]/4\pi$,
where $\mathbf{m}_1$ ($\mathbf{m}_2$) are the magnetic moments of the first (second) particle, $\hat{r}$ is in the direction of separation, and $\mu_0$ is the magnetic permeability of vacuum. Here, particle moments are along the magnetic field, $\hat{z}$, such that only the first term contributes. Regular thermal diffusion is negligible in our experiments due to the high P\'eclet number (${\rm Pe}\sim 100-300$, see End Matter), which quantifies advective motion relative to diffusion, ${\rm Pe} = v a/D$. Here, $a$ is the particle size, $D = 0.02\, \mu {\rm m}^2/s$ is the measured diffusion constant of the particles, and $v$ is the time averaged velocity of 300 particles.

The resulting expanding drop is very different from a drop of passive thermal particles --- for thermal particles, the boundary of a drop becomes diffused over time and has an unbounded tail (see Figs.~\ref{fig:brownian_motion_stuck_density}B and E in End Matter); here,  the profile is compact, and the boundary is sharp---the concentration is zero beyond a certain point [see Figs.~\ref{fig1}B--E and Supplemental Material video]. 
Particle positions were extracted  \cite{allan_2024_12708864} and the standard deviation (STD) of the positions was found to be ${\rm STD}(t)\propto R(t) \propto t^{\frac{1}{5}}$, where $R$ is the radius of the drop (Fig.~\ref{fig1}F).
Figure~\ref{fig1}E shows the rescaled radial average of the distribution as a function of distance, taken from four different experiments at three different times each (faded lines). The rigid dark blue line represents the average of all plots; the dots indicate the theoretical prediction.

\textbf{Numerical simulations.} To further validate the results, we numerically propagate the velocity using an eighth-order Runge-Kutta scheme with an adaptive time step. 
An initial random distribution of thousands of pointlike particles in a disk ($R_0 = 1$) interact through dipolar repulsion ($k=3$). 
Fig~\ref{fig1}G  is an overlay of three snapshots from a simulation.
By re-plotting the snapshots while rescaling the radius according to Eq.~\ref{self-similar ansatz}, they completely overlap; see Fig.~\ref{fig1}H. The self-similarity is further evident in Fig.~\ref{fig1}I, where the radially averaged rescaled densities fall on the same curve. 
Particles in both experiment and simulation show the formation of a triangular Wigner lattice\cite{PhysRev.46.1002}; see, e.g., the inset in Fig.~\ref{fig1}A. A measure of the local bond orientation parameter, $\Psi_6 = N^{-1}\sum_{i,j} e^{i 6 \theta_{ij}}$, gives $\Psi_6 \sim 0.9$ in simulations and $\Psi_6 \sim  0.7$ in  experiment. 
Since in our system, the distance between lattice sites varies, defects emerge to alleviate stress resulting from a mismatch in lattice spacing \cite{soni2018emergent,anderson2017theory,irvine2010pleats, Kelleher2015}. 

\textbf{Continuum model.} 
Let us examine the ensemble behavior, starting with pair interactions. For two particles repelled by $U \sim 1/r^k$ in the overdamped case, their separation grows as $r \propto t^{\frac{1}{k+2}}$. We show that this scaling also holds for the ensemble. At first glance, this scaling resembles thermal diffusion, where both a single Brownian particle and an ensemble spread as $\sqrt{t}$. However, Brownian particles are non-interacting, whereas power law repulsion involves many-body interactions. Such interactions could have altered the scaling, as indeed they do when the repulsion is not a power law \cite{ben2024compact}. 

\begin{figure}[htbp!]
\centering
\includegraphics[width = \linewidth]{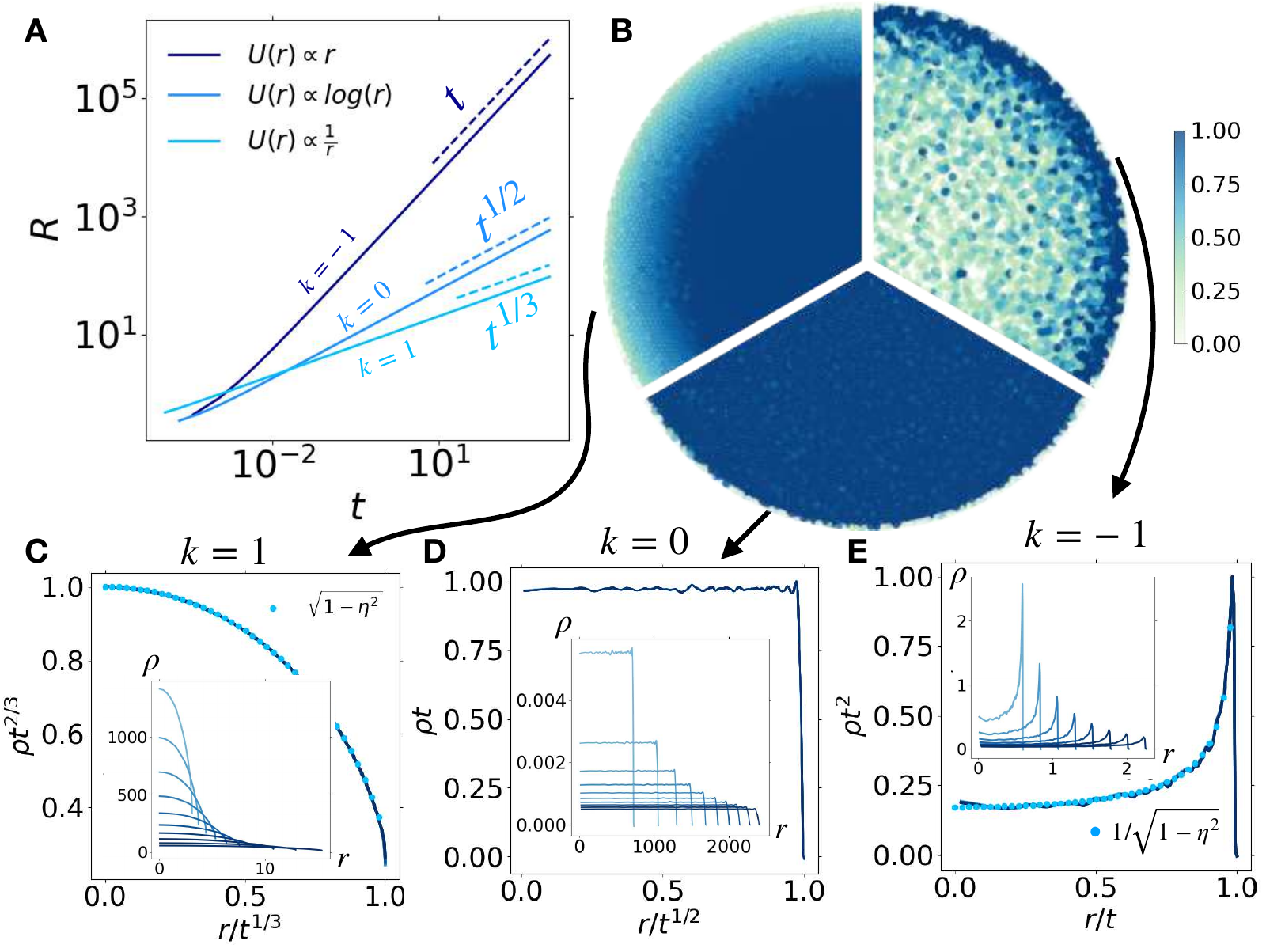}
\caption{Simulations in the long-range limit $k< d$. (A) Radius versus time for $k = (-1,0,1)$ in a simulation of 10000 particles, along with the theoretical scaling, $R(t) \propto t^{\frac{1}{k+2}}$, shown as a dashed line. (B) Snapshots of the distribution at the final time step, with color coding according to the density for k = 1 (top left), k = 0 (bottom), and k = -1 (top right). (C--E) Rescaled density according to $\rho t^{2\beta}$ as a function of the self-similar variable $\eta = r/t^{\beta}$. Bright blue dots give the theoretical predictions given in Eq.~\ref{f in thesis}. A single curve can be seen, but it is, in fact, ten plots at different times that collapse to a single curve. The insets show the separate, unscaled densities as a function of radius, with color going from light blue at early times to dark blue at later ones. }
\label{fig2}
\end{figure}
\begin{figure}[htbp!]
\centering
\includegraphics[width=0.95\linewidth]{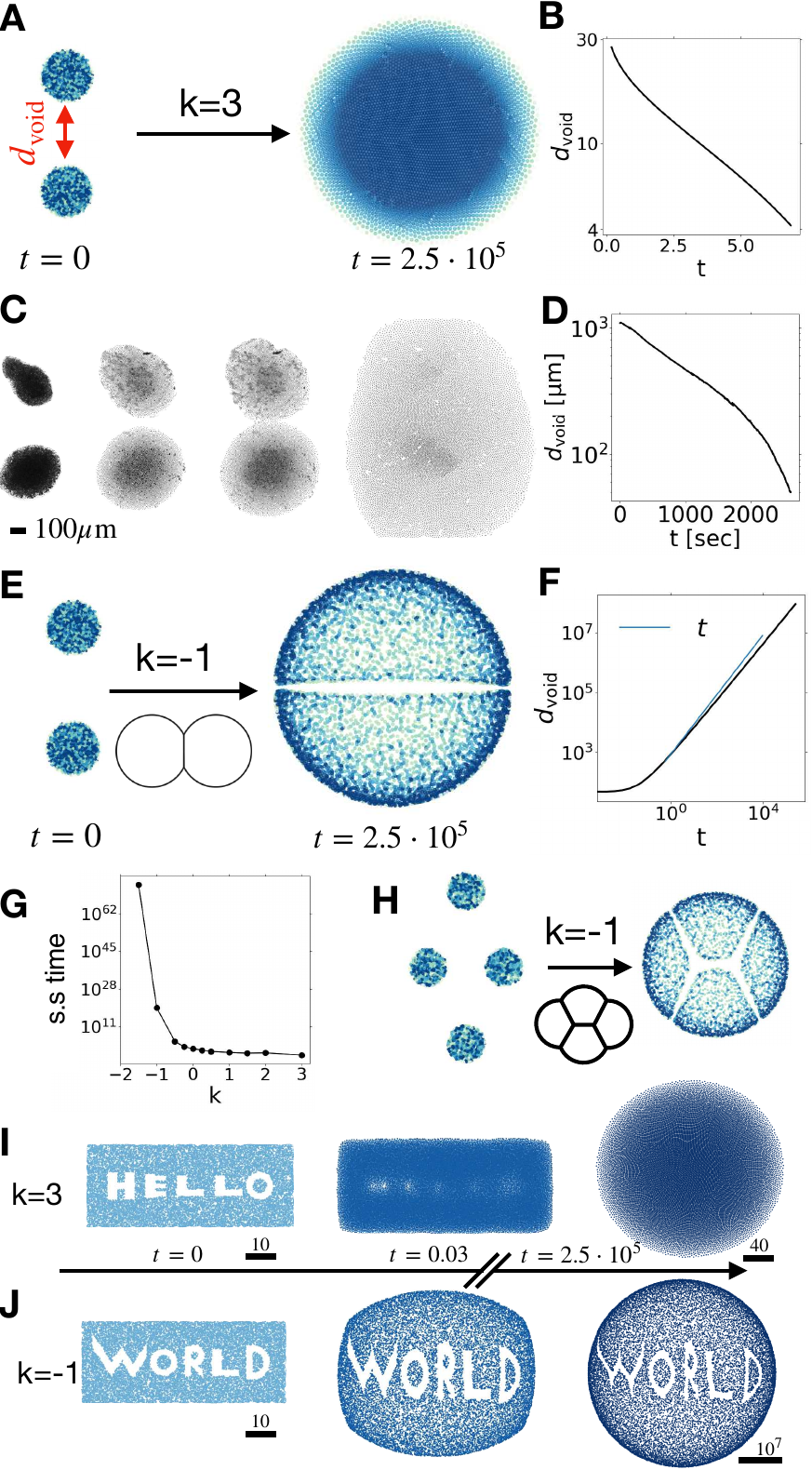} 
\caption{Memory retention in suspensions. (A, E) Collision of two suspensions of $N=2000$ particles each, for $k=3$ (in A) and $k=-1$ (in E). The final configuration is at $t=2.5\cdot 10^5$, showing an isotropic final state for $k=3$ and a memory of the initial condition for $k=-1$, with a visible corridor void of particles, $d_{\rm void}$. (B) and (F) plots of  $d_{\rm void}$ in time (B) For $k=3$, on a semi-log scale, and (F) for $k=-1$, on a log-log scale. The diagram in E shows two soap bubbles with a similar interface. (C) Experiment of two populations with $k=3$ reaching a merged state. Frames shown at ($t = 0, 10,200, 6600$\,s). (D) Plot of $d_{\rm void}$ in the experiment. (G) Time to reach an isotropic state, as a function of the power law exponent $k$. (H) Merging of 4 populations. (I,J) Memory of the initial conditions is quickly lost for $k=3$ (in I), and is maintained for $k=-1$ (in J). }
\label{fig3}
\end{figure}

To derive the self-similarity scaling, we will coarse-grain the dynamical equations. Considering $N$ particles that follow a power law potential, the continuity equation reads
$\frac{\partial \rho }{\partial t}+\nabla \cdot(\rho \textbf{v})=0$,
where $\textbf{v}$ is the velocity, and $\rho$, the density, defined by $\rho(\mathbf{r}) = \sum_{j}^{N} \delta(\boldsymbol{r_j} - \boldsymbol{r})$. Let us work with dimensionless variables such that $v\rightarrow v/v_0$, $ \mathbf{r} \rightarrow \mathbf{r}/R_0$, $R(t) \rightarrow R/R_0$, and $\rho \rightarrow \rho/\rho_0$, with $R_0$ being the initial radius of the suspension, and $\rho_0 = N/R_0^d$ is  the initial density. 
The number of particles is constant, giving a constraint: $N =\Omega_{d} \int_{0}^{\infty}\rho(r,t)r^{d-1}dr$,
where $\Omega_{d}$ is the spherical surface of the unit sphere in d-dimensions. Plugging the self-similar ansatz, $\rho(r,t)=At^{\gamma}f\left(\frac{Br}{t^{\beta}}\right)$, and requiring that the solution must not depend explicitly on time, gives $\gamma = -\beta d$.
The value of $\beta$ is derived from the potential $U(r)\propto 1/r^{k}$, by taking the velocity (Eq.~\ref{discrete velocity}) to the continuum limit. 
Using a non-dimensional variable $\eta = Br/t^{\beta}$, we arrive at $\beta = 1/(k+2)$, and an integral equation for $f$, the rescaled density,
\begin{equation}
\label{eqEta}
    \eta =\Gamma  \int_{0}^{\eta_{B}}\frac{\boldsymbol{\eta-\eta'}}{\left|\boldsymbol{\eta-\eta'} \right|^{k+2}}f(\eta')d^d\eta'\cdot \hat{r},
\end{equation}
where $\eta_{B}=(BR/t^{\beta})$ and we defined $\Gamma= B^{k+2-d}A(k+2)$, for more on Eq.~\ref{eqEta}; see End Matter. 
Note that $k$ can take values from $(-2,\infty]$. 
With $\beta$ and $\gamma$, the density of the suspension can be renormalized according to Eq.~\ref{self-similar ansatz}, i.e., $\rho(r,t)=At^{-d/(k+2)}f\left(\frac{Br}{t^{1/(k+2)}}\right)$.
We have shown this solution to be true in Fig.~\ref{fig1}, for both experiments and simulations with $k=3$. To further support this result, Fig.~\ref{fig2}A shows the radius as a function of time for simulations with $k = (-1,0,1)$ along the theoretical predictions $t^{1/(k+2)}$ \cite{krapivsky2024expansion}. 

We now describe the density profiles of the self-similar states. Figure~\ref{fig1} shows an example for a short-range power law $k=3$ in 2D, while Fig.~\ref{fig2}B presents simulations for long-range interactions ($k \leq d$) with $k = (-1,0,1)$. The angular averages over time, and rescaled densities via Eq.~\ref{self-similar ansatz} are presented as well. Notably, both the short-range case ($k=3$, Fig.~\ref{fig1}) and a long-range potential ($k=1$, Fig.~\ref{fig2}B, top left) lead to origin-centered densities, whereas other long-range cases ($ k=0,-1$, Fig.~\ref{fig2}B, bottom and top right) do not. 
We have tested simulations in 1D and 2D and can divide the profiles into three categories: (i) origin centered for $k>d-2$ with a hexagonal arrangement, (ii) constant profile at $k=d-2$ (apparently hyperuniform, see Fig. 5d in End Matter), and (iii) boundary centered for $k<d-2$.
Systems that interact via $k=d-2$ are termed a ``Coulomb gas" or a log-gas in 2D \cite{Minnhagen1987TheTC, leble2017large}. The boundary-centered density profile for strongly long ranged potentials ($k<d-2$), is referred to as the ``evaporation catastrophe" \cite{gallavotti1999statistical} and is attributed to the potential being too repulsive at infinity.
It is interesting to note (see Eq.~\ref{eqEta}) that the dynamical solution is similar to the equilibrium distribution when a harmonic potential confines the particles (see, e.g., \cite{Vieira2016Repuslive, Moreira2018PowerLaw, Agarwal2019Harmonic, krapivsky2024expansion}). For $d-2<k<d$, it has the same solution as for a nonlocal porous media equation \cite{biler2011barenblatt}. In End Matter, we briefly review the known results and give intuition for their derivation. The density profile is: 
\begin{align}
     \label{f in thesis}
     f(\eta)=
    \begin{cases}
     (1-\eta^{2})^{\frac{d}{k}}, & \text{if } k > d, \\
     (1-\eta^{2})^{\frac{k+2-d}{2}},  & \text{if } -2 < k < d.
    \end{cases}
\end{align}
Figures \ref{fig1}E and ~\ref{fig1}I show an excellent agreement between the experiment, the simulation, and the analytic solution for $k=3$ in 2D, and Fig.~\ref{fig2}C--E presents results for $k=(-1, 0, 1)$ in 2D. 

\textbf{Memory retention.}
We find that long-term memory of the initial distribution can be encoded in a system with strongly long-range repulsion when $k<d-2$. 
For that, we simulate collisions of two populations by initially placing two or more suspensions at a sufficient distance to prevent overlap, and measure $d_{void}$ defined as the distance between the two closest particles in each suspension. The merged isotropic state is universal (absorbing) and does not depend on initial conditions, but the relaxation rate differs dramatically with $k$: when $k
\geq d-2$, convergence to an isotropic state that is indistinguishable from a single ensemble is exponential, whereas for $k < d-2$, the system retains a power law memory of its initial state. Figure \ref{fig3} compares $k=3$, which quickly forms an isotropic state (in both simulations and experiments), with $k=-1$, where a particle‐free zone grows over time. Ultimately, even for $k=-1$, the suspensions merge into a single drop at very long times. We plot the time to reach a merged isotropic state in Fig.~\ref{fig3}G as a function of the power-law exponent $k$. This time approaches infinity as $k\rightarrow -2$. Collisions involving four populations are presented in Fig.~\ref{fig3}H, where an H-shaped particle-free zone is formed. The resulting shapes for both two and four populations are reminiscent of soap bubbles (see Fig.~\ref{fig3}E and H); however, in soap bubbles \cite{MorganFrankSoap}, the interaction is attractive, leading to the aggregation of particles at the boundary. Whereas in Fig.~\ref{fig3}E and H, the interaction is repulsive, and the boundary between the suspensions is particle-free. In Fig.~\ref{fig3}I and J, we further show the long-term memory retention for $k<d-2$. Fig.~\ref{fig3}I and J show an initial condition of particles with a word written by the absence of particles. Using a short-range potential with $k=3$, the word ``HELLO" quickly vanishes as the system evolves until all that is left is an isotropic suspension. Using a potential with $k=-1$, the written ``WORLD" remains as the system expands and evolves. Note also the change in scale between beginning and end.

\textbf{Discussion.}
We observed self-similar spreading of a dense ensemble of particles interacting by power law potentials, with a decay of $t^{-d/(k+2)}$. The density profile is compact, and takes the form of a semi-circle $(1-\eta^2)^\delta$, with (i) $\delta = d/k$ for short-ranged potentials ($k>d$) \cite{Agarwal2019Harmonic,Moreira2018PowerLaw} and (ii) $\delta = (k+2-d)/2$ for long-ranged potentials ($k<d$).
For strongly long-ranged interactions, $k<d-2$ ,\cite{biler2011barenblatt}, relaxation is anomalously slow ---  collisions between suspensions generate persistent particle-free regions,  paralleling slow relaxation of jammed and glassy materials \cite{liu2010jamming}.
Such interaction-driven memory effects may be relevant for phoresis in synthetic microswimmers ~\cite{Jin2018,Gompper2020} and communication in robotic swarms~\cite{Hamann2018,BenZion2023}.

\textit{Acknowledgments ---} The authors are grateful for discussions with Paul Krapibski, Kirone Mallick, Thomas Bickel, Yoav Lahini and Haim Diamant. This work was supported by an NSF-BSF grant number 2023624. 




\bibliographystyle{apsrev4-1}
\bibliography{Main_text}

\section{End Matter}
Here, we provide information on the derivation of the density distribution in the various limits, give more details on the experimental procedure and results for 5 ${\mu}$m ${\rm SiO}_2$ particles, show results from the 1D simulations, and include effects of diffusion. 

\begin{figure*}[htbp!]
        \includegraphics[width=0.8\linewidth]{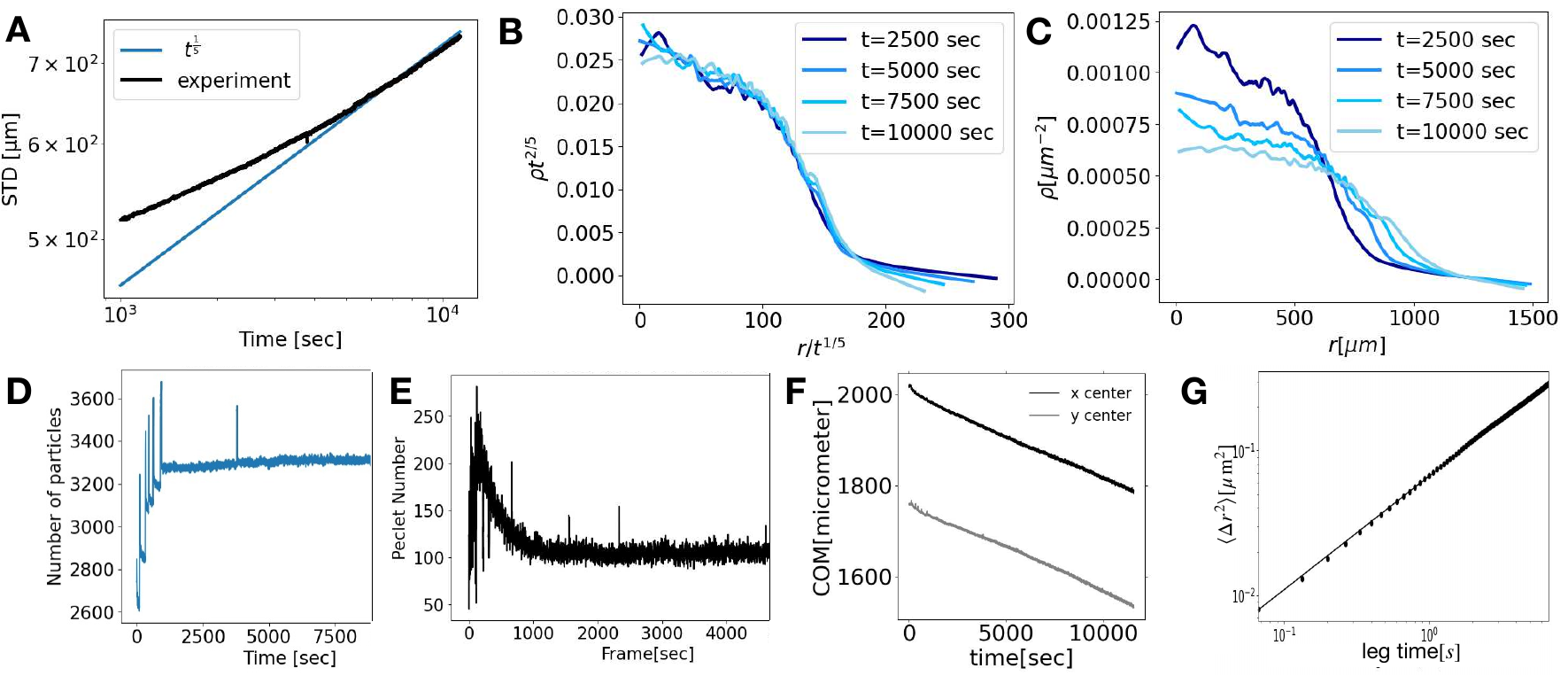}
        \caption{(A) STD of the ${\rm SiO}_2$ 5$\mu m$ particles as a function of time. (B) Normalized density of the ${\rm SiO}_2$ 5$\mu m$ particles according to the self-similar ansatz (Eq.1 in the main text). (C) Density of the ${\rm SiO}_2$ 5$\mu m$ particles at four different times. (D) Number of ${\rm SiO}_2$ particles as a function of time. (E) P\'eclet number of the PS 10$\mu$m particles throughout the experiment. (F) Drift of the center of mass. (G) Diffusion measurements. Ensemble average of the MSD versus leg time in a dilute suspension of PS particles with no magnetic field}
        \label{fig:Sio expriment}
\end{figure*}

\textbf{Derivation of Eq. (3):} For the isotropic initial conditions that we consider, the mass conservation equation becomes, 
$    \frac{\partial \rho}{\partial t} + \frac{1}{r^{d-1}}\frac{\partial (\rho r^{d-1}v)}{\partial r}=0$,
where $v$ is the radial velocity. Plugging the self-similar ansatz, Eq.~\ref{self-similar ansatz}, and requiring that the solution must not depend explicitly on time, gives $\gamma = -\beta d$.
Taking Eq.~\ref{discrete velocity} to the continuum limit gives 
$    \mathbf{v(r)}=  \int_{0}^{R}\frac{\mathbf{r-r'}}{\left|\mathbf{r-r'} \right|^{k+2}}\rho(\boldsymbol{r'})d^{d}r'$.
Using the non-dimensional variable $\eta=\frac{Br}{t^{\beta}}$, gives an integral equation for $f$, the rescaled density, which depends only on the self-similar variable $\eta$. This happens for
$\beta = 1/(k+2)$ and produces Eq.~\ref{eqEta}.

\textbf{Density for short-range interactions:}
For short-range interactions, i.e., in the limit $k>d$, we follow Ref.~\cite{Vieira2016Repuslive} but taking a slightly different route. 
In the short-range limit, faraway particles will have negligible contributions to the density at $\mathbf{\eta}$. Thus, we perform a Taylor expansion of the rescaled density $f(\boldsymbol{\eta}')$ around $\boldsymbol{\eta}$, $f(\boldsymbol{\eta'})=f(\boldsymbol{\eta}+\boldsymbol{s})\approx f(\boldsymbol{\eta})+\boldsymbol{s}\cdot \vec{\nabla} f(\boldsymbol{\eta}) + \cdots$,
where $\mathbf{s} = \boldsymbol{\eta'}-\boldsymbol{\eta}$. 
This results in, 
$\eta \approx -\Gamma \int_0^{\infty} \frac{\mathbf{s}}{s^{k+2}}\left[ f(\boldsymbol{\eta})+\mathbf{s} \cdot \nabla f(\boldsymbol{\eta})\right]d^ds \cdot \hat{r}$.
The first term vanishes after integration from symmetry considerations. The angular integral in the resulting term can be performed using the identity $\int \hat{s}\hat{s} d\Omega_d = \mathbf{I} \Omega_{d}/d$, on the surface of a sphere where $\hat{s} = \mathbf{s}/s$, and $\mathbf{I}$ is the identity tensor.
The radial part of the integral is of the form $\int^{\infty}_0 s^{d-1-k} ds$. The divergence at the lower boundary is resolved by taking it to be proportional to the mean distance between particles, $r_{\rm exc}=\alpha f^{-\frac{1}{d}}$ (see \cite{Vieira2016Repuslive}~\cite{Moreira2018PowerLaw}) since it is highly unlikely for two repulsive particles in a dissipative medium to collide. The integration gives
$   \eta = -\frac{k}{2d} {f^{k/d-1}}\frac{df}{d\eta}$,
where we defined $B^{k-d+2}=[k(k-d)]/[2d(k+2)\Omega_{d}  A \alpha^{d-k}]$, with $A=(N/\Omega_DI)B^d$ where $I=\int_{o}^{\infty}f(\eta)\eta^{d-1}d\eta$ is the constraint that the number of particles $N$ in the system is constant. This equation, in fact, also results from a nonlinear diffusion equation \cite{barenblatt1952porous, stone2002thinFilms,leal2007advanced, ben2024compact}.
The solution is, 
$f(\eta)=(1-\eta^{2})^{\frac{d}{k}}$.

\begin{figure*}[htbp!]
\centering
\includegraphics[width=0.95\linewidth]{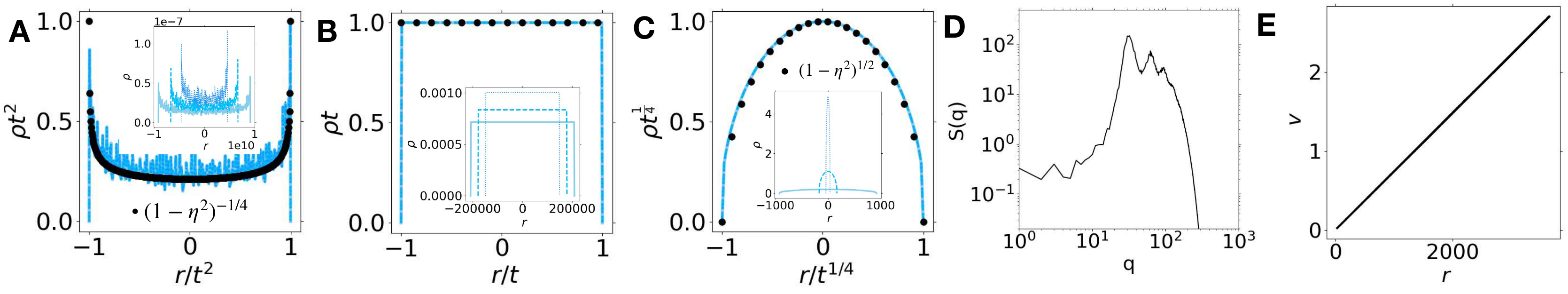}
\caption{\label{fig:density_for_k=-1_2_-15} Additional simulation results (A--C) Rescaled self similar density profiles from 1D simulations. (A) $k=-1.5$. (B) $k=-1$. (C) $k=2$. The inset shows the separate, unscaled
densities as a function of radius, with color going from
dark blue at early times to light blue at later ones. (D) Structure factor calculated from a 2D simulation with $k=0$ showing apparent hyperuniformity. (E) Velocity is linear with distance, measured from a 2D simulation with $k=0$.}
\end{figure*}

\textbf{Density for long-range interactions:}
The density profiles for 2D in Fig.~\ref{fig2}D and Fig.~\ref{fig:density_for_k=-1_2_-15}B for 1D, indicate that for $k=d-2$, the density is constant. A formal proof is given in Ref.~\cite{krapivsky2024expansion}. Here, let us give some intuition. For any solution to be self-similar, the velocity must be proportional to the distance from the center, $r$. That is $v \propto r$ (see Fig.\ref{fig:density_for_k=-1_2_-15}F for an example). Assuming a constant density in Eq.~\ref{eqEta}, the integral scales as $r^{-k-1+d}$. Equating this to $r$, gives $k=d-2$. Thus, for a potential of $U \sim 1/r^{d-2}$, the density profile is spatially constant. 
A solution for 1D is given in \cite{Agarwal2019Harmonic, krapivsky2024expansion}, and for D dimensions in \cite{biler2011barenblatt}. 

\textbf{Experimental procedure:}
The experiment is conducted as follows: Capillary tubes (0.2mm x 6mm x 50mm) are loaded with colloidal particles and concentrated using a bar magnet. Particles are magnetized by a uniform magnetic field of 3-5 mT,  generated using custom-built Helmholtz coils (see Fig.~\ref{fig1}). Contributions from induced dipole moments are negligible compared to the applied field. The sample is viewed using bright field illumination (Nikon Eclipse Ti2, $\times4$ magnification) and recorded at one frame every two seconds (Nikon-kinetix-m-c). With the magnetic field turned on, colloids mutually repel, and the suspension spreads (see Fig.~\ref{fig1}B--D).
The density of the suspension was calculated using Voronoi tessellation, taking the density of particle $i$ positioned at $\mathbf{r}$ to be the inverse of its Voronoi area, $\rho(\mathbf{r}_i) = 1/A_{\rm cell}^i$.

\textbf{Results for ${\rm SiO}_2$:}
Here, we present additional results for ${\rm SiO}_2$ particles of diameter 5$\mu$m. The ${\rm SiO}_2$ particles seemed to chain more than the PS particles in the same magnetic field (5 mT), we found that at a lower magnetic field (0.5 mT) the ${\rm SiO}_2$ particles chained less. Fig~\ref{fig:Sio expriment}A shows agreement between theory and practice, but only after a long time because of the weak magnetic field. The STD of the particles aligns with $R\propto t^\frac{1}{5}$ after about two hours. Fig~\ref{fig:Sio expriment}B shows the density of the suspension at a few different times, and Fig~\ref{fig:Sio expriment}C shows the rescaled density collapse to the same master curve given by Eq.~1.

\begin{figure*}[htbp!]
\centering
\includegraphics[width=0.95\linewidth]{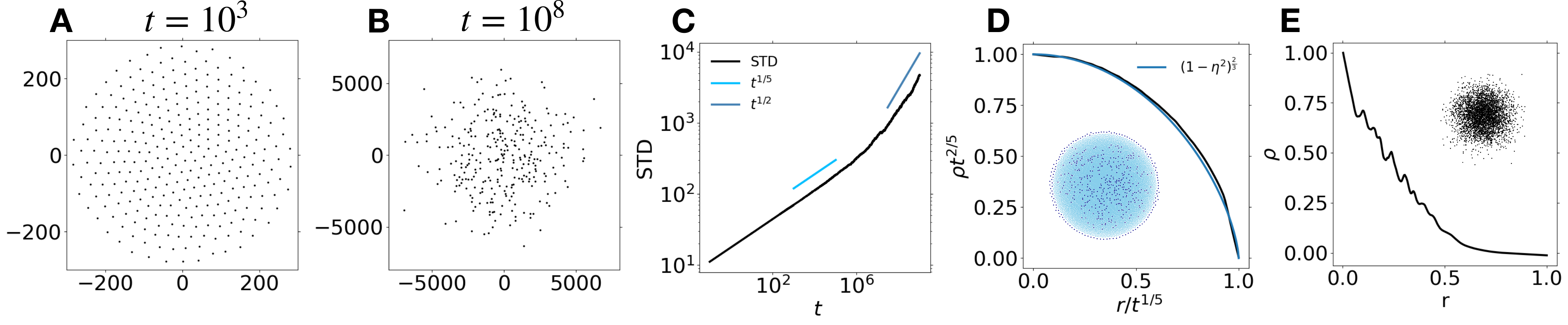}
\caption{\label{fig:brownian_motion_stuck_density} (A-C) Simulations of dipolar repulsion plus diffusion.  Snapshots of 300 particles taken at: (A) $t \sim 10^{3}$, where magnetic repulsion is dominant.  (B) $t \sim 10^{8}$, where diffusion is dominant. (C) The STD in the simulation, showing a transition from $t^{1/5}$ at short times to $t^{1/2}$ at long times. (D) Density of $10000$ particles interacting by $U\propto \frac{1}{r^3}$ where $5\%$ of particles have $\times 3$ the susceptibility than the other $95\%$.  (E) Density of $4000$ non-interacting Brownian particles. Insets show particles' spatial configuration corresponding to the density shown.}
\end{figure*}

\textbf{Chaining:}
The magnetic field of the Helmholtz coil induces magnetic dipoles in the particles in the $\hat{z}$ direction, resulting in positive potential energy and thus repulsion. However, due to different particle heights or non-uniformity of the field, magnetic dipoles can be induced with a component in the $\hat{\textbf{r}}$ direction \cite{C3SM00132F}. Turning the field off and on resets the induced dipoles and gives the particles time to distance each other, thus making some chains break apart. Fig~\ref{fig:Sio expriment}D shows the number of particles detected by the code in each frame. 
The number saturates after a few times we perform this procedure. In order to further reduce chaining, the particles were coated with $0.1\%w/v$ of Pluronic F-127. In order to conceptualize the chaining seen in experiments, we performed simulation with  $5\%$ of the particles with  $\times 3$ the susceptibility. Fig.~\ref{fig:brownian_motion_stuck_density}D shows an excellent agreement with theoretical predictions. 

\textbf{Drift:}
Drift in the sample can be observed in Fig.~\ref{fig:Sio expriment}F. 
The drift is relatively small, but became noticeable a couple of hours into the experiment. We reduced drift by degassing the suspension and a careful sample preparation.


\textbf{P\'eclet number:}
Diffusion measurement is presented in Fig.~\ref{fig:Sio expriment}G. Diffusion was measured for a dilute suspension of PS particles without an external magnetic field. Diffusion was extracted from the slope and is $D = 0.02 \mu{\rm m}^2/s$.
Fig~\ref{fig:Sio expriment}E, shows the P\'eclet number of the PS 10$\mu$m particles along the experiment, it is evident that $P_e\gg1$ throughout the experiment, which means that advection dominates.  To illustrate, a particle would diffuse its diameter over 21 min, and a comparable expansion, would take about a year. Fig.~\ref{fig:brownian_motion_stuck_density}C shows the STD of a simulation of $300$ particles interacting with dipolar magnetic repulsion and subjected to Brownian motion, $d \mathbf{x_{\rm tot} }= \mathbf{v} dt + \sqrt{2 dt D} \boldsymbol{\xi}$, where $\boldsymbol{\xi}$ is a random variable with a variance of 1, and we chose $D=0.02\mu{\rm m}^2/$s, and $v_0 l^{4} = 4.6 \cdot 10^4 \mu{\rm m}^{5}/{\rm s}$, as estimated experimentally. They show a transition from  spreading as $t^{\frac{1}{5}}$ (advection dominates) at short times to $t^{\frac{1}{2}}$ (Diffusion dominates) at long times. The transition occurs roughly at $t\sim 10^7 s$ which is roughly a year.

\textbf{1D simulation results:}
The initial configuration in 1D is that the particles are uniformly distributed on a line centered around the origin. 
The density in 1D is calculated in the same manner as in 2D, except for using the distance between particles instead of a Voronoi area. The density of each particle is defined as half the distance from each of its neighbors.
Fig.~\ref{fig:density_for_k=-1_2_-15}A Fig.~\ref{fig:density_for_k=-1_2_-15}B Shows the density profile for long-range potentials in 1D ($k=-1.5$ and $k=-1$ respectively) rescaled by Eq.~\ref{self-similar ansatz}, along with the analytical solution given by Eq.~\ref{f in thesis}, which shows excellent agreement between the two. Fig.~\ref{fig:density_for_k=-1_2_-15}C Shows the density of the short ranged potential ($k=2$) resacled by Eq.~\ref{self-similar ansatz}, along with the analytical solution given by Eq.~\ref{f in thesis}, which again shows excellent agreement between the two. The inset in each figure shows the unscaled density at three different times going from early times (dark blue) to later times (light blue).

\end{document}